# Amplification of whistler waves in a thin Keplerian disc embedded in an axisymmetric magnetic field

Yuri M. Shtemler[1], Michael Mond[1,2], and Günther Rüdiger[2]

[1]*Department of Mechanical Engineering, Ben-Gurion University of the Negev, P.O. Box 653, Beer-Sheva 84105, Israel*

[2]*AIP, Potsdam, Germany*

*20August2009*

**ABSTRACT**

The linear stability of thin Keplerian discs of weakly ionized polytropic plasma embedded in a mixed (toroidal-poloidal) axisymmetric magnetic field is studied. The effects of the central body on the equilibrium state of the discs are modeled by an axial total current and a dipole magnetic field. Studying the amplification of high-frequency waves shows that Keplerian discs can be destabilized due to combined effects of equilibrium density stratification and poloidal magnetic field. By considering a boundary value problem in the limit of strong Hall effect, it is shown that whistler waves which are free of direct influence of compressibility, rotation and gravity, are amplified along their trajectory while other modes of wave propagation do not change their amplitude.

# 1. INTRODUCTION

The dynamics of thin rotating gaseous discs under the influence of a magnetic field is of great importance to numerous astrophysical phenomena. While starting from Balbus & Hawley (1991) most of the research of rotating discs within an astrophysical context has focused on the magnetohydrodynamic (MHD) description of the magneto-rotational instability (MRI), the importance of the Hall effects to protoplanetary discs has been pointed out by Wardle (1999). Since then, such works as Balbus & Terquem (2001), Sano & Stone (2002), Salmeron & Wardle (2003), Salmeron & Wardle (2005), Desch (2004), Urpin & Rudiger (2005), Rudiger & Kitchatinov (2005) have shed light on the modification of the MRIs by the Hall effect.

In addition to modifying the MRIs, the Hall electromotive force gives rise to a new family of instabilities. As shown in Liverts & Mond (2004), and Kolberg et al. (2005), the Hall electric field combined with density stratification, in the presence of a constant vertical magnetic field, gives rise to a quasi-electrostatic (zero displacement current) mode that may become unstable for a relatively short density inhomogeneity length. That new family of instabilities is characterized by a corresponding dimensionless Hall parameter of the order of unity. Indeed, subsequent works by Shtemler et al. (2007) and Liverts et al. (2007) have demonstrated the relevance of the Hall instability to the dynamical development of rotating magnetized weakly ionized discs.

Although protoplanetary discs have been the subject of numerous investigations in search of possible instabilities, in most astrophysics related hydromagnetic stability studies, the models for the underlying equilibrium states are some times unrealistically simplified. Conventionally they are based either on the local approximation in both radial and axial coordinates, or on the model of an 'infinitely thick disc', an infinite cylinder that is uniform in the axial direction. Alternatively, a picture of an infinitesimally thin disc that lies in the disc midplane and whose properties vary with radius is employed. A more realistic model of thin rotating discs is adopted in Regev (1983), Ogilvie (1997), Kluzniak & Kita (2000) and Umurhan et al. (2006), and is based on an asymptotic approach in the small aspect ratio of the disc [see also Shtemler et al. (2007), where equilibrium thin discs embedded in a toroidal magnetic field and their stability were considered, and Shtemler et al. (2009), where equilibrium thin discs embedded in a toroidal-polidal axisymmetric magnetic field were investigated].

Furthermore, although the magnetic field configuration in real discs is largely unknown, observations and numerical simulations indicate that toroidal magnetic field may be of the same order of magnitude, or some times dominate the poloidal magnetic field [see e.g. Terquem & Papaloizou (1996); Papaloizou & Terquem (1997); Hawley & Krolik (2002); Proga (2003)]. In the initial stage of the disc rotation the poloidal field is commonly accepted as the dominant component. There are three kinds of possible axisymmetric equilibria of thin differentially rotating discs which can not be reduced from one to the other: (i) pure poloidal, (ii) pure toroidal and (iii) mixed poloidal-toroidal magnetic field (Shtemler et al. 2009). The first case, namely, steady discs with pure poloidal magnetic field was considered by Ogilvie (1997) within the classical MHD model for the equilibrium of a rotating thin disc, while Shtemler et al. (2007) have presented a general solution for the second class of thin disc equilibria and their stability for the Hall MHD case.

In the present study the dynamical response to high frequency perturbations of the third class of the Hall equilibrium in mixed poloidal-toroidal magnetic field is investigated. The equilibrium solutions are approximated by a double asymptotic expansion, namely in the small aspect ratio $\varepsilon$ of the thin disc, and in the small parameter $\delta_H$ that is inversely proportional to the Hall parameter [Shtemler et al. (2009)]. It has been demonstrated that depending on the effect of the central body two classes of Hall MHD disc equilibrium may occur: discs of small ($R_d \sim \delta_H^0$) and large ($R_d \sim \delta_H^{-m}$, $m>0$) radius. As the local approximation for investigating the stability of such equilibrium configurations is invalid under such general non-uniform conditions, an alternative approach is adopted. Thus, instead of the traditional eigen-value problem that is customarily solved, namely finding the complex frequencies for all given real wave numbers, an alternative strategy is applied in which the spatial amplification of waves of a given real frequency is calculated by an appropriate WKB-like method. Unlike the eigen-value problem for stratified thin Hall MHD discs, such an approach is mathematically consistent with non uniform equilibrium conditions, and its physical meaning is transparent as it describes the spatial amplification of perturbations that are generated at some specified location of the disc. It is thus the aim of the current work to carry out a systematic investigation of the non-local instabilities that may arise in an equilibrium of Hall MHD rotating discs of weakly ionized polytropic plasma, embedded in a mixed poloidal-toroidal axisymmetric magnetic fields.

In order to carry out that scheme, a sound understanding of the properties of the various modes of wave propagation in high-Hall-parameter plasmas is necessary. Indeed, the correspondence between the characteristic waves in a MHD plasma and their counterparts in a Hall MHD (HMHD) plasmas with increasing Hall effect was discussed in Swanson (2003) and in Hameiri et al. (2005). Thus, Hameiri et al. (2005) have obtained a full set of dispersive HMHD waves in the limit of short wavelengths in a general non-rotating 3D system under strong Hall effects. They found dramatic changes in the nature of the classical MHD waves (Alfvén, fast-and slow-magnetosonic waves) when the Hall electric field increases. Three types of waves are found, one of which moves with the sound velocity, while the two others move with super- and sub-sonic velocities. Following the accepted convention of referring to the middle (sound) wave as the Alfvén wave, Hameiri et al. (2005) call these waves as Alfvén, super-Alfvénic and sub-Alfvénic waves, respectively. They further demonstrate that the MHD fast magnetosonic wave becomes incompressible and an essentially magnetic wave that is reduced in the shortwave limit to a whistler wave. The slow magnetosonic wave is incompressible and free from magnetic effects. Most dramatically, the incompressible MHD shear Alfvén wave, turns for strong Hall effects into a compressible acoustic wave with negligible perturbations of the magnetic field. Furthermore, the fast and slow MHD magnetosonic waves become in the large Hall parameter limit incompressible, the fast wave becomes essentially magnetic while the slow wave becomes mainly of hydrodynamic nature.

Bearing in mind the properties of the various wave propagation modes, a wave amplification model is applied in the high-frequency limit to discs of large radius, one of the two classes found in the steady-state analysis, namely, of small and large radius discs (Shtemler et al, 2009). It is demonstrated in the present study that Keplerian discs may become unstable due to the combined effects of the Hall electromotive force, and the equilibrium density stratification that gives rise to whistler waves. It is found that the whistler waves which are free from the direct influence of compressibility, rotation, gravity as well as of the toroidal magnetic field, may be significantly amplified along their trajectory. In particular, following the classification of Hameiri et al. (2005), the instability of the super-Alfvénic (whistler) wave corresponds to Hall branch of instabilities [Shtemler et al. (2007), Liverts et al. (2007)], while the middle-Alfvén (sound) waves that corresponds to the Hall modified MRI branch are stable in thin disc approximation. Thus, the present results are quite distinct from the results of the stability analysis for thin Keplerian discs in pure toroidal magnetic fields as discussed in Shtemler et al. (2007), where compressibility, rotation and gravity effects do determine the Hall instability.

It should finally be noted that the finite thickness of the disc as well as its internal structure play a crucial role in determining its stability. Thus, as calculations that model the disc as an infinite cylinder predict the existence of strong MRI Hall modified instability (see e.g., Sano & Stone 2002 in the limit of weakly-compressible plasma), recent works point out the stabilizing effects of the discs' finite thickness in the limit of small thickness to radius ratio [Coppi & Keyes (2003), Liverts & Mond (2009)]. It is thus envisaged that propagating back and forth within the disc due to multiple reflections from the boundaries, the amplified waves may lead to sustainable absolute-like growing perturbations [Liverts & Mond (2009)]. Thus, an accurate description of the structure of thin equilibrium discs is crucial for stability study. Indeed, in that limit, as will be shown in the subsequent sections, it is the whistler mode that gets amplified rather than the appropriate high Hall parameter analog of the Alfvén waves.

The paper is organized as follows. The dimensionless governing equations are presented in Section 2. In section 3 the results are summarized for the Hall-MHD equilibrium configuration of a rotating thin disc subjected to a gravitational potential of a central body and mixed magnetic fields. A model of amplification of perturbations in thin equilibrium Hall discs is presented in Section 4. Application of the high-frequency approximation to whistler waves in the equilibrium discs of large radius with strong magnetic dipoles is in Section 5. Summary and discussion are given in Section 6. Appendix contains a short description of possible equilibrium solutions for thin Hall discs developed by an additional asymptotic expansion in small inverse Hall parameter.

## 2. THE PHYSICAL MODEL FOR HALL DISCS

The stability of radially and axially stratified thin partially ionized rotating plasma discs threaded by a mixed toroidal and poloidal magnetic field is considered. Viscosity and radiation effects are ignored.

It is convenient transform all the physical variables to non-dimensional quantities by using the following characteristic values [Shtemler et al. (2007), (2009)]:

$$V_* = \Omega_* r_*, \quad t_* = 1/\Omega_*, \quad \Phi_* = V_*^2, \quad m_* = m_i, \quad n_* = n_n,$$

$$P_* = K(m_* n_*)^\gamma, \quad j_* = \frac{c}{4\pi} \frac{B_*}{r_*}, \quad E_* = \frac{V_* B_*}{c}.$$

Here $\Omega_* = (GM_c/r_*^3)^{1/2}$ is the Keplerian angular velocity of the fluid at the characteristic radius $r_*$ that belongs to the Keplerian portion of the disc ($r_*$ may be considered as the inner boundary of the Keplerian portion of the disc); $G$ is the gravitational constant; $M_c$ is the total mass of the central object; $\Phi_*$ is the is the characteristic value of the gravitational potential; $m_*$ and $n_*$ are the characteristic mass and number density; $m_i = Zm_p$ ($Z = 1$ for simplicity), $m_i$ and $m_p$ are the ion and proton masses; $K$ is the dimensional constant in the polytropic law $P = Kn^\gamma$, $\gamma$ is the polytrophic coefficient, ($\gamma = 5/3$ in the adiabatic case); $c$ is the light speed. The characteristic values of the electric current density and electric field, $j_*$ and $E_*$, have been chosen consistently with Maxwell's equations. The characteristic dimensional magnetic field $B_*$ will be specified below. The resulting dimensionless dynamical equations are:

$$\frac{D(Pn^{-\gamma})}{Dt} = 0, \tag{1}$$

$$n\frac{D\mathbf{V}}{Dt} = -\frac{1}{M_S^2}\nabla P + \frac{1}{\beta M_S^2}\mathbf{j}\times\mathbf{B} - n\nabla\Phi, \qquad \Phi(r,z) = -\frac{1}{(r^2+z^2)^{1/2}}, \tag{2}$$

$$\frac{\partial n}{\partial t} + \nabla\cdot(n\mathbf{V}) = 0, \tag{3}$$

$$\frac{\partial \mathbf{B}}{\partial t} + \nabla\times\mathbf{E} = 0, \qquad \nabla\cdot\mathbf{B} = 0, \tag{4}$$

$$\mathbf{E} = -\mathbf{V}\times\mathbf{B} + \Pi_H\frac{\mathbf{j}\times\mathbf{B}}{n}, \qquad \mathbf{j} = \nabla\times\mathbf{B}. \tag{5}$$

Standard cylindrical coordinates $\{r,\theta,z\}$ are adopted throughout the paper with the associated unit vectors $\{\mathbf{i}_r,\mathbf{i}_\theta,\mathbf{i}_z\}$; $\mathbf{V}$ is the plasma velocity; $t$ is time; $D/Dt = \partial/\partial t + (\mathbf{V}\cdot\nabla)$ is the material derivative; $\Phi$ is the gravitational potential due to the central object. The electric field $\mathbf{E}$ is described by the generalized Ohm's law which is derived from the momentum equation for the electron fluid by neglecting the electron inertia and pressure as well as the effect of ambipolar diffusion [Pandey & Wardle (2008)]; $\mathbf{B}$ is the magnetic field, $\mathbf{j}$ is the current density; $P = P_e + P_i + P_n$ is the total plasma pressure; $P_l = n_l T_l$ are the partial species pressures ($l = e; i; n$); $T = T_e = T_i = T_n$ is the plasma temperature; the masses of electron, ion and neutrals are as follows $m_e \ll m_i \approx m_n$; subscripts $e$, $i$ and $n$ denote the electrons, ions and neutrals, respectively.

Since the plasma is assumed to be quasi-neutral and partially ionized with small ionization degree $\alpha$, strongly coupled ions and neutrals and the number density of plasma $n$

$$n_e \approx n_i \approx \alpha n_n, \quad n \approx n_n, \quad \alpha = n_e/(n_e + n_n) \ll 1, \quad V_i \sim V_n. \tag{6}$$

Note that a preferred direction is tacitly defined here, namely, the positive direction of the $z$ axis is chosen according to positive Keplerian rotation. The dimensionless coefficients $M_S$, $\beta$, and $\Pi_H$ are the Mach number, plasma beta and Hall coefficient, respectively:

$$M_S = \frac{V_*}{C_{S*}}, \quad \beta = 4\pi \frac{P_*}{B_*^2}, \quad \Pi_H = \frac{\Omega_i}{\Omega_*}\left(\frac{l_i}{r_*}\right)^2 \equiv \frac{B_* c}{4\pi e \alpha n_* \Omega_* r_*^2}, \tag{7}$$

$C_{S*} = \sqrt{P_*/(m_* n_*)}$ is the characteristic sound velocity; $l_i = C_{S*}/\omega_{pi}$ and $\Omega_i = eB_*/(m_* c)$ are the inertial length and the Larmor frequency of ions, respectively; $\omega_{pi} = \sqrt{4\pi e^2 \alpha n_*/m_*}$ is the plasma frequency of the ions; $e$ is the electron charge. The increasing importance of the Hall term for weakly ionized plasmas is apparent as $\Pi_H$ is inversely proportional to the ionization degree $\alpha$.

In real protoplanetary discs, the electron density is determined by ionization versus recombination, and the ionization fraction may vary significantly in space. In such cases, rate equations that describe the ionization degree should be employed. However, in the present study the ionization fraction is assumed to be constant. This simplification is made to avoid the widely uncertain physics of ionization and recombination. Nevertheless, such approximation allows one to roughly estimate the real characteristics of the system [Shtemler et al. (2007)].

A common property of thin Keplerian discs is their highly compressible motion with large Mach numbers $M_S$ [Frank, King, & Rairle (2002)]. Furthermore, the characteristic effective semi-thickness $H_* = H(r_*)$ of the disc ($H = H(r)$ is the local disc half thickness) is such that the disc aspect ratio $\varepsilon$ equals the inverse Mach number:

$$\frac{1}{M_S} = \varepsilon = \frac{H_*}{r_*} \lesssim 1. \qquad (8)$$

The smallness of $\varepsilon$ means that dimensionless axial coordinate is also small, i.e. $z/r_* \sim \varepsilon$ ($|z| \leq H$), and consequently the following rescaled values may be introduced in order to further apply the asymptotic expansions in $\varepsilon$ [Shtemler et al. (2007), (2009)]:

$$\zeta = \frac{z}{\varepsilon} \sim \varepsilon^0, \quad h = \frac{H}{\varepsilon} \sim \varepsilon^0, \quad \pi_H = \frac{\Pi_H}{\varepsilon} \sim \varepsilon^0. \qquad (9)$$

## 3 HALL EQUILIBRIUM SOLUTIONS FOR A MIXED MAGNETIC FIELD IN THIN DISC LIMIT

Summarizing and slightly modifying the results presented in Shtemler et al. (2009) it is first noted that the solution of the steady-state equilibrium problem is obtained within the Keplerian portion of magnetized discs by asymptotic solution of Eqs. (1)-(5) in small aspect ratio $\varepsilon$ (similar to Regev (1983), Ogilvie (1997), Kluzniak & Kita (2000), Umurhan et al. (2006)). This yields for the gravitational potential:

$$\Phi(r,z) = \Phi^{(0)}(r) + \varepsilon^2 \Phi^{(2)}(r,\zeta) + O(\varepsilon^4), \quad \Phi^{(0)}(r) = -\frac{1}{r}, \quad \Phi^{(2)}(r,\zeta) = \frac{\zeta^2}{2r^3}, \quad r > 1 \gg \varepsilon, \quad \zeta \sim \varepsilon^0. \qquad (10)$$

To leading order in $\varepsilon$ the toroidal velocity $V_\theta$ is described by the Keplerian law, which follows from the leading order radial component of the momentum equation, while $V_r$, $V_z$ are of higher order in $\varepsilon$ as in Ogilvie (1997), and their input to the stability analysis is negligibly small. Thus,

$$\mathbf{V} \approx V_\theta \mathbf{i}_\theta + O(\varepsilon), \quad V_\theta = \Omega(r) r \sim \varepsilon^0, V_r \sim \varepsilon, \quad V_z \sim \varepsilon^2, (\Omega(r) = r^{-3/2}). \qquad (11)$$

The divergence-free ($\nabla \cdot \mathbf{B} = 0$) axisymmetric equilibrium magnetic field $\mathbf{B}$ has the following toroidal-poloidal decomposition with components, $B_\theta$ and $\mathbf{B}_p$:

$$\mathbf{B} = B_\theta \mathbf{i}_\theta + \frac{1}{r} \hat{\nabla} \Psi \times \mathbf{i}_\theta, \quad \Psi = \varepsilon \psi, \quad B_r = -\frac{1}{r} \frac{\partial \psi}{\partial \zeta}, \quad B_z = \varepsilon \frac{1}{r} \frac{\partial \psi}{\partial r}, \quad (\hat{\nabla} = \mathbf{i}_r \frac{\partial}{\partial r} + \frac{1}{\varepsilon} \mathbf{i}_z \frac{\partial}{\partial \zeta}), \qquad (12)$$

where $\boldsymbol{B}_p = B_r \boldsymbol{i}_r + B_z \boldsymbol{i}_z$ is expressed through the flux function $\Psi$ which is scaled with $\varepsilon$ in such a way that the resulting radial component of the magnetic field is of the order of the toroidal one ($\sim \varepsilon^0$), while the axial magnetic field $B_z$ is of the order of $\varepsilon$. In addition, Farady's law yields:

$$rB_\theta(r,\zeta) = I(\psi) \tag{13}$$

where $I(\psi)$ is the total current flowing through a circular cross section defined by the given $r$ and $\zeta$. Furthermore, Ampere's law is written as follows:

$$\boldsymbol{E} \equiv \varepsilon \hat{\nabla} \varphi = -\boldsymbol{V} \times \boldsymbol{B} + \varepsilon \pi_H \frac{\boldsymbol{j} \times \boldsymbol{B}}{n}, \quad \varphi = \varphi(\psi) \tag{14}$$

The above relations result to leading order in $\varepsilon$ in the following nonlinear differential equation for $\psi$ in $\zeta$, which depends parametrically on the radial coordinate $r$:

$$\frac{\partial^2 \psi}{\partial \zeta^2} = -I(\psi)\frac{dI}{d\psi} - \frac{1}{\pi_H} N(r,\zeta)[\Omega(r) + \frac{d\varphi}{d\psi}]. \tag{15}$$

Here $N = r^2 n_v$ is the moment of inertia density within the Keplerian disc. Equation (15) is the Grad-Shafranov (GS) equation for the special case of a thin disc in Hall MHD equilibrium. To avoid solving a coupled problem inside and outside the Keplerian disc, in which the edge of the disc $\zeta = h(r)$ is determined self-consistently, the following assumption is adopted:

$$h(r) \equiv const = 1. \tag{16}$$

Consequently, the differential equation (15) must be complemented by appropriate boundary conditions at the disc edge as well as by symmetry conditions at the midplane:

$$n = 0, \ \psi = \Gamma(r) \text{ at } \zeta = h(r) \equiv 1, \ \frac{\partial \psi}{\partial \zeta} = 0 \text{ at } \zeta = 0. \tag{17}$$

Due to lack of direct observational information on magnetic field distribution along the disc edge the following idealized flux function at the edge $\zeta = 1$ for the equilibrium Keplerian discs is examined:

$$\Gamma(r) = -M_d / r, \tag{18}$$

where a dipolar magnetic field of strength $M_d$ is introduced that qualitatively reflects the effect of the central body on the Keplerian portion of the disc, while the primordial magnetic field is assumed to produce a much smaller influence on the equilibrium than the dipole like magnetic field. The sign of the dipole strength $M_d$ determines the orientation of the magnetic dipole axis with respect to direction of the angular velocity of Keplerian rotation.

Employing Ohm's law in order to integrate the axial momentum equation (8), and using Ampere's law result in the following explicit relation for the number density to leading order in $\varepsilon$:

$$n(r,\zeta) = \nu \left\{ \frac{2}{\beta_H}[\varphi(\psi) - \varphi(\Gamma)] + \frac{2\Omega(r)}{\beta_H}[\psi(r,\zeta) - \Gamma(r)] + \frac{1-\zeta^2}{r^3} \right\}^{\frac{1}{\gamma-1}}, \quad (\nu = (\frac{\gamma-1}{2\gamma})^{\frac{1}{\gamma-1}}). \quad (19)$$

Here $\psi(r,\zeta)$ is a solution of the GS equation (15) with the corresponding boundary function $\Gamma(r)$. The arbitrary functions $\varphi=\varphi(\psi)$ and $I=I(\psi)$ in Eqs. (15), (19) reflect the uncertainty of the steady-state equilibrium that as assumed here describes the unsteady solution at long times. In general the corresponding initial value problem shifts that uncertainty to the arbitrariness of the initial data. The resulting equilibrium relations, rewritten in terms of the $\nu$-scaled number density, $n_\nu = n/\nu$, and moment of inertia density, $N_\nu \equiv N/\nu$, reveal that they contain the Hall parameter $\pi_H$ in the two following combinations: as the ratio $\delta_H = \nu/\pi_H$ in the GS equation (15), and as a factor with the plasma beta $\beta_H = \beta\pi_H$ in Eq. (19). Since both the plasma beta as well as the Hall parameter are independent physical parameters, a scaled Hall plasma beta parameter $\beta_H$ and a scaled inverse Hall parameter $\delta_H$ are naturally introduced instead of $\beta$ and $\pi_H$:

$$\beta_H = \beta\pi_H, \quad \delta_H = \frac{\nu}{\pi_H}, \quad (\beta_H \delta_H \equiv \beta\nu). \quad (20)$$

In further analysis $\beta_H$ is assumed to be of the order of or larger than unity, $\beta_H \gtrsim 1$. An assumption $\delta_H = \nu/\pi_H \lesssim 0.1$ is valid for Hall parameter $\pi_H \gtrsim 1$ ($\nu \approx 0.1$ for $\gamma = 5/3$), and is violated in the classical

MHD limit for vanishing Hall parameter $\pi_H$. The above general problem for thin equilibrium discs is further discussed in Appendix for some trial functions, $I(\psi)$ and $\varphi(\psi)$, and boundary function $\Gamma(r)$. The central body effects are included into the model adopted through the arbitrary functions $I(\psi)$ and $\Gamma(r)$. Several partial limits in small $\delta_H$ are presented in the Appendix.

## 4 AMPLIFICATION OF PERTURBATIONS IN THIN DISCS

Perturbations in equilibrium axisymmetric thin Keplerian discs embedded in a mixed poloidal-toroidal magnetic field can now be investigated. As in the case of the equilibrium problem, the perturbed problem is studied within thin disc approximation. We start by linearizing the Hall MHD equations (1) - (5) about the equilibrium solution. For that purpose, the perturbed variables are given by

$$f(r,z,t) = F(r,\zeta) + F'(r,\zeta,t). \tag{21}$$

Here $f$ stands for any of the physical variables, $F$ and $F'$ denote the equilibrium and perturbation values, respectively, such that

$$|\frac{1}{F'}\frac{\partial F'}{\partial \zeta}| \sim |\frac{1}{F'}\frac{\partial F'}{\partial r}| \gg |\frac{1}{F}\frac{\partial F}{\partial \zeta}| \sim |\frac{1}{F}\frac{\partial F}{\partial r}|. \tag{22}$$

Similar to the steady-state equilibrium relations (12), the divergence-free axisymmetric magnetic field for the unsteady perturbations $\boldsymbol{B}'$ may also be decomposed into toroidal and poloidal components, $B'_\theta \sim \varepsilon^0$ and $\boldsymbol{B}'_p \sim \varepsilon^0$:

$$\boldsymbol{B}' = B'_\theta \boldsymbol{i}_\theta + \frac{1}{r}\hat{\nabla}\Psi' \times \boldsymbol{i}_\theta,\ \Psi' = \varepsilon\psi',\ B'_r = -\frac{1}{r}\frac{\partial \psi'}{\partial \zeta} \sim \varepsilon^0,\ B'_z = \varepsilon\frac{1}{r}\frac{\partial \psi'}{\partial r} \sim \varepsilon,\ (\hat{\nabla} = \boldsymbol{i}_r\frac{\partial}{\partial r} + \boldsymbol{i}_z\frac{1}{\varepsilon}\frac{\partial}{\partial \zeta}), \tag{23}$$

It is convenient at this stage to rescale the physical variables with the small parameters $\nu$ and $\varepsilon$:

$$n' = \nu n_\nu^{(0)'},\qquad P' = \nu P_\nu^{(0)'}, \tag{24}$$

$$\boldsymbol{V'} = \{\varepsilon^2 V_r^{(2)\prime}, \varepsilon^2 V_\theta^{(2)\prime}, \varepsilon V_z^{(1)\prime}\}, \qquad \boldsymbol{B'} = \{B_r^{(0)\prime}, B_\theta^{(0)\prime}, \varepsilon B_z^{(1)\prime}\}, \qquad \boldsymbol{E'} = \{\varepsilon E_r^{(1)\prime}, \varepsilon E_\theta^{(1)\prime}, E_z^{(0)\prime}\}, \qquad (25)$$

such that the scaled variables are now of order unity in both small parameters. The superscript denotes the order in $\varepsilon$, while the subscript $\nu$ denotes the values that are scaled with $\nu$. The scaling with $\nu$ in (24) allows the rewriting of the perturbation problem in terms of the effective parameters of the equilibrium problem $\beta_H$ and $\delta_H$ with no dependence on $\nu$. The scaling with $\varepsilon$ in (25) reduces the resulting problem to a regular form that is uniformly valid in the limit of $\varepsilon \ll 1$. Thus, to leading order in $\varepsilon$, Eqs. (1)-(5) yield the following set of linearized equations:

$$\frac{\partial P_\nu^{(0)\prime}}{\partial t} = c_S^2 \frac{\partial n_\nu^{(0)\prime}}{\partial t}, \qquad (c_S^2 = \gamma n^{(0)\gamma-1} \equiv \frac{\gamma-1}{2} n_\nu^{(0)\gamma-1}), \qquad (26)$$

$$n_\nu^{(0)} \frac{\partial V_z^{(1)\prime}}{\partial t} = -\frac{1}{\beta_H \delta_H} \frac{\partial (B_\theta^{(0)} B_\theta^{(0)\prime} + B_r^{(0)} B_r^{(0)\prime})}{\partial \zeta} - \left[\frac{\partial P_\nu^{(0)\prime}}{\partial \zeta} + n_\nu^{(0)\prime} \frac{\partial \Phi^{(2)}}{\partial \zeta}\right], \qquad (27)$$

$$\frac{\partial n_\nu^{(0)\prime}}{\partial t} + \frac{\partial n_\nu^{(0)} V_z^{(1)\prime}}{\partial \zeta} = 0, \qquad (28)$$

$$\frac{1}{r} \frac{\partial \psi^{(0)\prime}}{\partial t} + E_\theta^{(1)\prime} = 0, \qquad (29)$$

$$\frac{\partial B_\theta^{(0)\prime}}{\partial t} + \frac{\partial E_r^{(1)\prime}}{\partial \zeta} - \frac{\partial E_z^{(0)\prime}}{\partial r} = 0, \qquad (30)$$

where

$$B_r^{(0)\prime} = -\frac{1}{r} \frac{\partial \psi^{(0)\prime}}{\partial \zeta}, \qquad B_z^{(1)\prime} = \frac{1}{r} \frac{\partial \psi^{(0)\prime}}{\partial r}, \qquad (31)$$

$$E_\theta^{(1)\prime} = -V_z^{(1)\prime} B_r^{(0)} + \frac{1}{\delta_H n_\nu^{(0)}} \left[\frac{1}{r} \frac{dI}{d\psi^{(0)}} \left(\frac{\partial \psi^{(0)}}{\partial \zeta} B_z^{(1)\prime} + \frac{\partial \psi^{(0)}}{\partial r} B_r^{(0)\prime}\right) + B_z^{(1)} \frac{\partial B_\theta^{(0)\prime}}{\partial \zeta} + B_r^{(0)} \frac{1}{r} \frac{\partial r B_\theta^{(0)\prime}}{\partial r}\right],$$

$$E_r^{(1)\prime} = -r\Omega B_z^{(1)\prime} + B_\theta^{(0)} V_z^{(1)\prime} - (\Omega + \frac{d\varphi}{d\psi}) r B_z^{(1)} \frac{n_\nu^{(0)\prime}}{n_\nu^{(0)}} + \frac{1}{\delta_H n_\nu^{(0)}} \left[\frac{\partial B_r^{(0)}}{\partial \zeta} B_z^{(1)\prime} + B_z^{(1)} \frac{\partial B_r^{(0)\prime}}{\partial \zeta} - \frac{1}{r^2} \frac{\partial (r B_\theta^{(0)})(r B_\theta^{(0)\prime})}{\partial r}\right],$$

$$E_z^{(0)}{'} = r\Omega B_r^{(0)}{'} + (\Omega + \frac{d\varphi}{d\psi})rB_r^{(0)} \frac{n_v^{(0)}{'}}{n_v^{(0)}} - \frac{1}{\delta_H n_v^{(0)}} \frac{\partial(B_\theta^{(0)} B_\theta^{(0)}{'} + B_r^{(0)} B_r^{(0)}{'})}{\partial \zeta}. \qquad (32)$$

The equations for $V_r^{(2)}{'}$ and $V_\theta^{(2)}{'}$ that describe the perturbed motion in the disc plane are decoupled from the above system, and are governed by the following equations:

$$n_v^{(0)}[\frac{\partial V_r^{(2)}{'}}{\partial t} - 2\Omega V_\theta^{(2)}{'}] = -\frac{\partial P_v^{(0)}{'}}{\partial r} + \frac{1}{\beta_H \delta_H} \{\frac{\partial B_r^{(0)}}{\partial \zeta} B_z^{(1)}{'} + B_z^{(1)} \frac{\partial B_r^{(0)}{'}}{\partial \zeta} - \frac{1}{r^2} \frac{\partial (rB_\theta^{(0)})(rB_\theta^{(0)}{'})}{\partial r}\}, \qquad (33)$$

$$n_v^{(0)}[\frac{\partial V_\theta^{(2)}{'}}{\partial t} + \frac{1}{2}\Omega V_r^{(2)}{'}] = \frac{1}{\beta_H \delta_H}[\frac{\partial B_\theta^{(0)}}{\partial \zeta} B_z^{(1)}{'} + \frac{1}{r}\frac{\partial rB_\theta^{(0)}}{\partial r} B_r^{(0)}{'} + B_z^{(1)} \frac{\partial B_\theta^{(0)}{'}}{\partial \zeta} + B_r^{(0)} \frac{1}{r} \frac{\partial rB_\theta^{(0)}{'}}{\partial r}]. \qquad (34)$$

If not mentioned otherwise the superscript denoting the orders in small aspect ratio $\varepsilon$ are dropped from now in both equilibrium and perturbation variables in order to simplify the notations.

## 5. HIGH-FREQUENCY APPROXIMATION FOR FUNDAMENTAL WAVES IN THIN DISCS

For further simplicity and analytical accessibility, magnetic configurations with equilibrium stream function that depends only on the radial variable are investigated:

$$\psi(r,\zeta) \equiv \psi(r), \qquad B_r \equiv 0, \quad rB_z(r) = d\psi(r)/dr, \quad rB_\theta(r) = I(\psi(r)). \qquad (35)$$

Such equilibrium solutions approximately describe large-radius thin discs with strong magnetic dipoles (Appendix). Other magnetic field configurations presented in Appendix may be considered by a similar way, however (35) is the simplest one which demonstrates the patterns of typical perturbation growth.

### 5.1 Whistler waves

As the steady-state of the disc is highly spatially inhomogeneous, the customary eigen-value approach is replaced here by an appropriate boundary value problem. According to the latter, perturbations are generated at some location within the disc or at its boundary with a given real frequency $\omega$=const. The ensuing propagation of those perturbations into the disc is described by a phase function $\chi(r,\zeta)$ such that at each point within the disc and at each time the perturbation is given by (see a discussion in Section 6):

$$F'(r,\zeta,t) = \tilde{F}\exp[-i\omega t + i\omega_H^{1/2}\chi(r,\zeta)], \qquad (36)$$

where the frequency $\omega_H = \delta_H \omega$ is naturally scaled similar to (20). For facilitating an analytic solution the high frequency limit, $\omega_H \gg 1$, is considered.

Substituting (35) - (36) into Eqs. (26)-(32) and omitting the superscript that denotes the order in small aspect ratio $\varepsilon$ yields with the aid of (11), (12) and (23), (25):

$$\tilde{P}_v - c_S^2 \tilde{n}_v = 0, \qquad (c_S^2 = \frac{\gamma-1}{2}n_v^{\gamma-1}), \qquad (37)$$

$$-i\omega n_v \tilde{V}_z + \frac{ik_\zeta}{r^2 \beta_H \delta_H} I(\psi) r\tilde{B}_\theta = -(ik_\zeta + \frac{\partial}{\partial \zeta})\tilde{P}_v - \frac{\partial \Delta\Phi}{\partial \zeta}\tilde{n}_v - \frac{I(\psi)}{r^2 \beta_H \delta_H}\frac{dr\tilde{B}_\theta}{d\zeta}, \qquad (\Delta\Phi = \frac{\zeta^2}{2r^3}), \qquad (38)$$

$$-i\omega \tilde{n}_v + ik_\zeta n_v \tilde{V}_z = -\frac{\partial n_v \tilde{V}_z}{\partial \zeta}, \qquad (39)$$

$$-\omega\tilde{\psi} + \frac{B_z}{\delta_H n_v}k_\zeta r\tilde{B}_\theta = \frac{B_z}{\delta_H n_v}[\frac{dI(\psi)}{d\psi}k_\zeta \tilde{\psi} + i\frac{\partial r\tilde{B}_\theta}{\partial \zeta}], \qquad (B_z(r) = \frac{1}{r}\frac{d\psi}{dr}), \qquad (40)$$

$$-\omega r\tilde{B}_\theta + \frac{B_z}{\delta_H n_v}k_\zeta^3 \tilde{\psi} = i\frac{B_z}{\delta_H n_v}k_\zeta[3k_\zeta\frac{\partial \tilde{\psi}}{\partial \zeta} + 3\frac{\partial k_\zeta}{\partial \zeta}\tilde{\psi} - \frac{k_\zeta}{N_v}\frac{\partial N_v}{\partial \zeta}\tilde{\psi}] - \frac{I(\psi)r}{\delta_H N_v^2}[\frac{\partial N_v}{\partial \zeta}k_r - \frac{\partial N_v}{\partial r}k_\zeta]r\tilde{B}_\theta. \qquad (41)$$

Here $k_r$ and $k_z$ are the local wave numbers

$$k_r(r,\zeta) = \omega_H^{1/2}\frac{\partial \chi(r,\zeta)}{\partial r}, \qquad k_\zeta(r,\zeta) = \omega_H^{1/2}\frac{\partial \chi(r,\zeta)}{\partial \zeta}, \qquad (\frac{\partial k_r}{\partial \zeta} \equiv \frac{\partial k_\zeta}{\partial r}). \qquad (42)$$

The following relations have also been utilized:

$$r\tilde{B}_r = -\frac{\partial \tilde{\psi}}{\partial \zeta} - \tilde{\psi}ik_\zeta, \quad r\tilde{B}_z = \frac{\partial \tilde{\psi}}{\partial r} + \tilde{\psi}ik_r. \qquad (43)$$

Equations (37)-(41) have been derived by inserting ansatz (36) into the appropriate equations (26)-(32) and by accounting for the two highest order terms in $\omega_H$. The following assumed orders of the main variables are self-consistent with the system of equations system (37)-(41):

$$r\tilde{B}_\theta \sim \omega_H^{1/2},\ \tilde{\psi} \sim \tilde{V}_z \sim \omega_H^0,\ \tilde{n}_v/\delta_H \sim \tilde{P}_v/\delta_H \sim \omega_H^{-1/2},\ k_r \sim k_\zeta \sim \omega_H^{1/2},\ \chi \sim \omega_H^0. \tag{44}$$

Estimates (44) may be justified a' posteriori after the calculations of the eigen-values and eigen-functions. Small terms of the order of or less than $\delta_H$ are additionally neglected in Eqs. (37)-(41).

The solution of the linearized equations for the amplitude functions $\tilde{F}$ and eigen-values $k_r$, $k_\zeta$ is obtained by asymptotic expansions in high frequency. Therefore, in this section and below, when it does not lead to misunderstanding, the superscripts in the expansions for $\tilde{F}$ and $k_r$, $k_\zeta$ in $\omega_H \gg 1$ will be indicated, while, as have been stated before, superscripts corresponding to the orders in small aspect ratio $\varepsilon$ are dropped.

It is first noted that due to the high-frequency scaling (44), Eqs. (40) and (41) for $\tilde{\psi}$ and $r\tilde{B}_\theta$ are decoupled from the rest of the system, and the following asymptotic expansions are adopted:

$$k_r = k_r^{(1)}\omega_H^{1/2} + k_r^{(0)}\omega_H^0 + \ldots,\quad k_\zeta = k_\zeta^{(1)}\omega_H^{1/2} + k_\zeta^{(0)}\omega_H^0 + \ldots,\quad \omega_H^{1/2}\chi = \chi^{(1)}\omega_H^{1/2} + \chi^{(0)}\omega_H^0 + \ldots,$$

$$\tilde{\psi} = \tilde{\psi}^{(0)}\omega_H^0 + \tilde{\psi}^{(-1)}\omega_H^{-1/2} + \ldots,\quad r\tilde{B}_\theta = r\tilde{B}_\theta^{(1)}\omega_H^{1/2} + r\tilde{B}_\theta^{(0)}\omega_H^0 + \ldots, \tag{45}$$

where

$$k_r^{(j)}(r,\zeta) = \frac{\partial \chi^{(j)}}{\partial r},\ k_\zeta^{(j)}(r,\zeta) = \frac{\partial \chi^{(j)}}{\partial \zeta},\qquad (j=0,1). \tag{46}$$

Inserting now (45) into (40) and (41) and neglecting the right-hand sides to leading order in $\omega_H^{1/2}$ yield the following eigen-value problem:

$$-\omega_H^0 \tilde{\psi}^{(0)} + \frac{B_z}{n_v}k_\zeta^{(1)} r\tilde{B}_\theta^{(1)} = 0, \tag{47}$$

$$-\omega_H^0 r\widetilde{B}_\theta^{(1)} + \frac{B_z}{n_\nu} k_\zeta^{(1)3}\widetilde{\psi}^{(0)} = 0 \quad . \tag{48}$$

The resulting eigen-value $k_\zeta^{(1)}$ is given by

$$k_\zeta^{(1)} = \frac{\partial \chi^{(1)}(r,\zeta)}{\partial \zeta} = \pm\sqrt{\frac{\omega_H^0 n_\nu}{|B_z(r)|}}, \tag{49}$$

which in turn yields the phase and radial component of the wave vector $\chi^{(1)}$ and $k_r^{(1)}$

$$\chi^{(1)}(r,\zeta) = \pm\sqrt{\frac{\omega_H^0}{|B_z(r)|}} \int_0^\zeta \sqrt{n_\nu(r,\zeta)} d\zeta , \quad k_r^{(1)} = \frac{\partial \chi^{(1)}(r,\zeta)}{\partial r}.$$

Although the scaled frequency is identically equal to unity ($\omega_H^0 \equiv 1$), the notation $\omega_H^0$ is left in order to get a physically transparent meaning of (49). Thus, evidently that Eq. (49), resolved formally with respect to $\omega_H^0$, is the conventional dispersion relation for a whistler mode [see e.g. Krall & Trivelpiece (1973)]. The arbitrary function of $r$ is set to zero, when $\chi^{(1)}(r,\zeta)$ is calculated in (49) by integration of $k_\zeta^{(1)}(r,\zeta)$ in $\zeta$, i.e. $\chi^{(1)}(r,\zeta) = 0$ at the mid-plane $\zeta = 0$. Thus, according to the Appendix, the functions that determine the phase function, i.e. the pure-hydrodynamic number density $n_\nu = n_\nu(r,\zeta)$, and the axial magnetic field $rB_z = d\psi(r)/dr$ within large discs are given by (superscripts in Eqs. (A7)-(A8) are omitted):

$$n_\nu(r,\zeta) = n_{\nu m}(r)\left(1-\zeta^2\right)^{\frac{1}{\gamma-1}}, \quad \psi(r,\zeta) \equiv -M_d r^{-1}, \quad B_z = M_d r^{-3}, \; n_{\nu m}(r) = r^{-\frac{3}{\gamma-1}}, \tag{50}$$

where $n_{\nu m} = n_\nu(r,0)$ is the number density at the mid-plane of the disc.

Finally, in order to find the eigen-functions, it is noted that since one of the linearly dependent relations (47) and (48) may be dropped, and one of the amplitudes either $\widetilde{\psi}^{(0)}$ or $r\widetilde{B}_\theta^{(1)}$, e.g. $\widetilde{\psi}^{(0)}$, may be set equals to an arbitrary constant, one has by using (43) and (47)

$$\widetilde{\psi}^{(0)} = 1, \quad r\widetilde{B}_r^{(1)} = -ik_\zeta, \; r\widetilde{B}_z^{(1)} = ik_r, \quad r\widetilde{B}_\theta^{(1)} = sign(M_d)k_\zeta^{(1)}(r,\zeta). \tag{51}$$

Moving now to the next order in $\omega_H^{1/2}$, expansions (45) are substituted into Eqs. (40)-(41) which after some algebra and using Eqs. (49) and (51) yield:

$$\frac{B_z}{n_v} k_\zeta^{(1)} r\widetilde{B}_\theta^{(0)} - \omega_H^0 \widetilde{\psi}^{(-1)} = \frac{|B_z| k_\zeta^{(1)}}{n_v} k_\zeta^{(0)} + \frac{B_z k_\zeta^{(1)}}{n_v} \frac{dI(\psi)}{d\psi} + i\frac{|B_z|}{n_v} \frac{\partial k_\zeta^{(1)}}{\partial \zeta}, \tag{52}$$

$$\frac{B_z}{n_v} k_\zeta^{(1)} r\widetilde{B}_\theta^{(0)} - \omega_H^0 \widetilde{\psi}^{(-1)} = \frac{3\omega_H^0}{k_\zeta^{(1)}} k_\zeta^{(0)} + \frac{I(\psi)r}{N_v^2}\left[\frac{\partial N_v}{\partial \zeta} k_r^{(1)} - \frac{\partial N_v}{\partial r} k_\zeta^{(1)}\right] - ir^2 |B_z| \left[3N_v^{-1} \frac{\partial k_\zeta^{(1)}}{\partial \zeta} + k_\zeta^{(1)} \frac{\partial N_v^{-1}}{\partial \zeta}\right]. \tag{53}$$

Since the right-hand side of the inhomogeneous system (52)-(53) coincides with the eigen-value problem (47) - (49) for $k_\zeta^{(1)}$, the value of $k_\zeta^{(0)} = \text{Re}\{k_\zeta^{(0)}\} + i\,\text{Im}\{k_\zeta^{(0)}\}$ is determined by the solvability condition of the system (52)-(53), which is:

$$k_\zeta^{(0)} \equiv \frac{\partial \chi^{(0)}(r,\zeta)}{\partial \zeta} = \left\{\frac{I(\psi)}{2|B_z(r)|rN_v}\left[\frac{\partial N_v}{\partial r} - \frac{\partial N_v}{\partial \zeta}\frac{k_r^{(1)}}{k_\zeta^{(1)}}\right] + \frac{\text{sign}(B_z)}{2}\frac{dI(\psi)}{d\psi}\right\} + i\frac{1}{2n_v}\frac{\partial n_v}{\partial \zeta}. \tag{54}$$

Finally, assuming that $\text{Im}\{\chi^{(0)}\}=0$ at $\zeta = 0$, and integrating (54) yield the following imagine parts of the wave vector and phase of the zero order in frequency:

$$\text{Im}\{k_\zeta^{(0)}\} = \frac{\partial \text{Im}\{\chi^{(0)}(\zeta)\}}{\partial \zeta} = \frac{1}{2n_v}\frac{\partial n_v}{\partial \zeta}, \quad \text{Im}\{\chi^{(0)}(\zeta)\} = \frac{1}{2}\ln\frac{n_v(r,\zeta)}{n_{vm}(r)}, \quad \text{Im}\{k_r^{(0)}\} = \frac{\partial \text{Im}\{\chi^{(0)}(\zeta)\}}{\partial r} \equiv 0. \tag{55}$$

Since the imagine part of the phase is the logarithm of the square root of the number density ratio, the non-exponential amplification of the whistler wave characteristics $F'$ which represents either $\psi'$ or $rB_\theta'$ takes the following form:

$$F'(r,\zeta,t) = \widetilde{F}(r,\zeta)\exp[-i\omega t + i\omega_H^{1/2}\chi] \sim A(\zeta)\widetilde{f}(r,\zeta)\exp[-i\omega t], \tag{56}$$

where $\widetilde{f}(r,\zeta)$ is a bounded function on $\zeta$, which can be easily calculated for all eigen-functions, while the amplification factor $A(\zeta)$ that tends to infinity at the disc edges is:

$$A(\zeta) = \exp(-\text{Im}\{\chi^{(0)}(\zeta)\}) \equiv \sqrt{\frac{n_{vm}}{n_v}} = \frac{1}{\sqrt{(1-\zeta^2)^{\frac{1}{\gamma-1}}}}, \qquad (n_{vm}(r) = r^{-\frac{3}{\gamma-1}}). \tag{57}$$

Thus, compressibility, rotation and gravity effects have no influence on a whistler wave amplification,

which occurs in Hall plasmas due to density stratification. In particular, in the approximation adopted, the amplification factor is independent of the choice of arbitrary equilibrium trial functions $I(\psi)$, $\varphi(\psi)$ and boundary function $\Gamma(r)$ (see Appendix). Furthermore, in the disc axial direction the whistler wave amplification $A(\zeta)$ is a power function of the distance from the horizontal edges of the disc that tends to infinity at the disc edges ($\zeta = \pm 1$), while the amplification factor is independent of the radial distance. In addition it is clear that the wave number of the whistler waves is singular as well as the amplification factor at the horizontal disc's edges, $\zeta = \pm 1$, where the density vanishes, $n=0$

It is finally noted that whistlers are electromagnetic in nature, which means that they perturb primarily the electromagnetic fields, and are incompressible to leading order. Indeed, Eqs. (40)-(41) that are decoupled from Farady's law result in zero perturbations of both velocity and pressure to leading order. This is in contrast to the fast magnetosonic wave to which the whistlers transition in the small Hall parameter limit (the MHD limit), who are highly compressible with equal perturbations of the magnetic field and all velocity components [see also Hameier et al.(2005)].

Once $\widetilde{\psi}^{(0)}$ and $r\widetilde{B}_\theta^{(1)}$ are known, the rest of relations (37)-(39) of the stability problem reaffirm the validity of assumptions (44) and determine the eigen-functions for the higher order perturbed density, pressure and velocity.

**5.2 On sound and inertial waves**

For wave numbers other than those found for the whistler waves as proper solutions of the eigen-value problem described by the sub-system (40) - (41), the latter has only a trivial solution for the magnetic field, namely:

$r\widetilde{B}_\theta = 0$, $\widetilde{\psi} = 0$. (58)

As a result, equations (37)-(39) that govern the hydrodynamic perturbations in the plane normal to the disc are decoupled and describe sound waves. In order to see that, it is assumed that the perturbations are characterized now by a given real frequency $\omega$ and a phase function which is scaled as $\omega^{1/2}\chi(r,\zeta)$ (different from (36) for the whistler waves):

$$F'(r,\zeta,t) = \widetilde{F}(r,\zeta)\exp[-i\omega t + i\omega^{1/2}\chi(r,\zeta)]. \tag{59}$$

The following expansions in high-frequency approximation are now applied which are self-consistent with Eqs. (26)-(32) in the absence of the magnetic field:

$$\widetilde{n}_\nu = \widetilde{n}^{(0)}\omega^0 + \widetilde{n}^{(-1)}\omega^{-1} + ..., \quad \widetilde{P}_\nu = \widetilde{P}^{(0)}\omega^0 + \widetilde{P}^{(-1)}\omega^{-1} + ...,$$

$$\widetilde{V}_z = \widetilde{V}_z^{(0)}\omega^0 + \widetilde{V}_z^{(-1)}\omega^{-1} + ..., \qquad k_\zeta = k_\zeta^{(0)}\omega^0 + k_\zeta^{(-1)}\omega^{-1} + .... \tag{60}$$

Substituting Eqs. (59)-(60) into Eqs. (26)-(32) yields to leading order in $\omega$:

$$-i\omega^0 n\widetilde{V}_z^{(0)} + ik_\zeta^{(1)} c_S^2 \widetilde{n}^{(0)} = 0, \tag{61}$$

$$-i\omega^0 \widetilde{n}^{(0)} + ik_\zeta^{(1)} n\widetilde{V}_z^{(0)} = 0. \tag{62}$$

System (61) - (62) has a nontrivial solution only for those eigen-values $k_\zeta^{(1)}(r,\zeta)$ that are given by:

$$k_\zeta^{(1)}(r,\zeta) = \pm\omega^0 / c_S(r,\zeta), \quad (c_S^2 = \gamma n^{\gamma-1}). \tag{63}$$

Here the notation $\omega^0$ is used for the scaled frequency that is identically equals to unity ($\omega^0 \equiv 1$), in order to get a physically transparent interpretation of (63) which evidently describes sound waves. Comparing asymptotic expansions for $k_\zeta$, note that amplification/absorption factor of sound waves in (60) is much less than the amplification factor for whistler waves in (46).

Interestingly, this mode is the reincarnation of the Alfv'en wave in the limit of strong Hall effect [Swanson (2003), Hameiri et al. (2005)]. In contrast to the incompressible classical MHD Alfv'en wave that carries mainly perturbations in both the magnetic field as well as in the fluid velocity, its HHMD counterpart is of compressible hydrodynamic nature, such that the electromagnetic perturbations are much smaller than the fluid dynamical ones. Furthermore, as the Alfv'en waves are responsible for the MRI, it is plausible to conclude that MRIs are not excited or at least not amplified within the thin disc, small iverse-Hall parameter approximation. This result is in accord with recent eigen-value calculations [Coppi & Keyes (2003), Liverts & Mond (2008)].

Since as indicated above sound, and whistler waves are, respectively, the HMHD counterparts of Alfv'enic, and fast magnetosonic waves in classical MHD it can be conjectured that inertial waves which are described by equations (33)-(34) which are neglected, are counterparts of slow magnetosonic waves.

## 5.3 Numeric modeling of whistler waves in high frequency approximation

The characteristics of large equilibrium discs that are characterized by a strong magnetic dipole (see Appendix) are presented in Fig. 1.

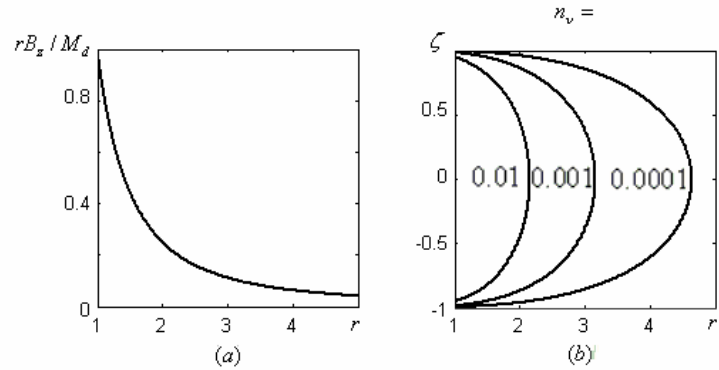

Figure 1. Characteristics of Keplerian equilibrium discs of large radius.
Small deviations from the reference total current, $(I-1) \sim \delta_H$; a strong magnetic-dipole strength, $M_d \sim \delta_H^0$.
(*a*) Equilibrium magnetic field $rB_z(r)/M_d \equiv d\psi/dr = 1/r^3$; $B_r \equiv 0$; $rB_\theta = I(\psi) \approx 1$;
(*b*) Contours of constant equilibrium number density, $n_\nu = 0.01, 0.001, 0.0001$, ($\gamma \approx 5/3$).
$$n_\nu(r,\zeta) \cong n_{\nu m}(r)(1-\zeta^2)^{\frac{1}{\gamma-1}}, \quad n_{\nu m}(r) = r^{-\frac{3}{\gamma-1}}.$$

In Figure 2 the scaled phase and wave numbers of whistler waves are depicted to leading order in high-frequency. Due to (36) the constant phase value $\chi = const$ depicts the front location of the whistler wave at the times $t = \chi^{(1)} \delta_H / \omega_H^{1/2}$. According to Fig. 2, the perturbation is generated at the mid-plane and the wave front propagates from the mid-plane to the horizontal edges $\zeta = \pm 1$ of the Keplerian region $r \geq 1$. Note that the values of phase $\chi^{(1)}$ in Fig. 2 correspond to time instants $t = \chi^{(1)} \delta_H / \omega_H^{1/2} \ll 1$, i.e. much less than the dimensionless Keplerian time $t_K \sim 1$. So that the whistler waves fastly expand through almost the whole Keplerian portion of the disc, and may undergo multiple reflections from the horizontal edges during a single Keplerian rotation time.

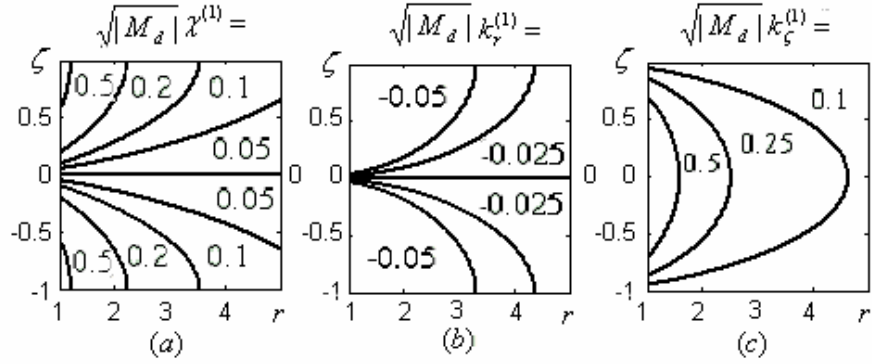

Figure 2. Whistler-wave characteristics to the leading-order in high-frequency approximation. Contours of constant phase (a) $\chi^{(1)}$; and of constant wave numbers as well as related components of the magnetic field (b) $k_r^{(1)} = -ir\widetilde{B}_z^{(1)}$; (c) $k_\zeta^{(1)} = r\widetilde{B}_\theta^{(1)} / sign(M_d) = ir\widetilde{B}_r^{(1)}$, ($\gamma \approx 5/3$).

Amplification factor of whistler waves depicted in Fig. 3 demonstrates according to Eq. (59) that whistler waves are indeed strongly amplified in the near edge zones.

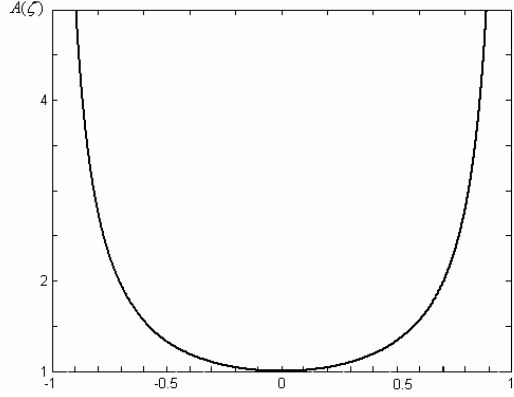

Figure 3. Whistler-wave amplification factor $A(\zeta)$ to zero-order in high-frequency approximation.

$$A(\zeta) = \exp(-\chi^{(0)}(\zeta)) = 1/\sqrt{\frac{n_v(r,\zeta)}{n_{vm}(r)}}, \qquad \frac{n_v(r,\zeta)}{n_{vm}(r)} = (1-\zeta^2)^{\frac{1}{\gamma-1}}, \qquad n_{vm}(r) = r^{-\frac{3}{\gamma-1}}, \ (\gamma \approx 5/3).$$

## 6. SUMMARY AND DISCUSSION

Linear HMHD theory of axisymmetric perturbations growth in equilibrium astrophysical discs is developed for weakly ionized polytropic plasmas. Thin equilibrium Keplerian discs that are embedded in an arbitrary poloidal-toroidal axisymmetric magnetic field are investigated.

As the equilibrium state of thin discs is highly inhomogeneous, the guiding strategy of the current investigation has been to follow the inward propagation of perturbations that are generated in the mid-plane with a given real frequency. The phase function of such perturbations describes its subsequent propagation such that the gradient of the phase function represents the generalized local wave vector. Finding the latter as the eigen-value of the linearized system, its imaginary part signifies the amplification or absorption of the waves along their trajectories.

First, general thin disc approximation relations were developed for perturbations in an equilibrium axisymmetric magnetic field that satisfies the GS equation and the divergent-free condition. It is found that the relations which describe the perturbed hydrodynamic motion in the disc plane are decoupled from the full system. In the present study the equilibrium stream function is assumed for simplicity to depend only on the radial variable. It is shown by asymptotic expansions in small inverse Hall parameter that such steady-state solutions describe the equilibrium state of thin discs of large-radius with a strong magnetic dipole that is provided by the central body. In the high-frequency limit the perturbed problem for thin discs is reduced to three sub-problems which correspond to the fundamental modes:

(i) fast-velocity whistler magnetic mode with no hydrodynamic motion and compressibility effects; (ii) sound mode describing a non-magnetic middle-velocity compressible motion developed in the axial direction; (iii) a pure hydrodynamic incompressible inertial mode that describes the motion in the disc plane. As was already mentioned in the Introduction, sound, whistler and inertial waves are, respectively, the HMHD counterparts of Alfv'enic, fast and slow magnetosonic waves in conventional MHD [Swanson (2003) and in Hameiri et al. (2005)]. It has been shown here that within the approximation adopted, those three fundamental modes are of different orders in the high frequency limit. Thus, whistler waves are characterized by the largest-order amplification factor as compared with sound and inertial waves and thus may be investigated independently. The picture that has been emerged is that of high frequency, fast propagating whistler waves whose amplitude is significantly amplified along its trajectory due to the strong vertical density stratifications. As the Alfv'en waves are responsible for the MRI, it may be concluded, in accord with recent eigen-value calculations [Coppi & Keyes (2003), Liverts & Mond (2008)] that MRIs are not excited or at least not amplified within the thin disc, high Hall parameter approximation.

It is immediately clear due to spatial inhomogeneity of the equilibrium discs that taking the traditional path of assuming a real wave vector, and obtaining the solutions for the complex eigen-frequencies will lead to a dead end. In the high-wave-number approximation this occurs due to combined effect of the perturbed Hall electric current and spatial inhomogeneity of the equilibrium solution. To leading order in wave-number, the non-local effects of the equilibrium are negligible for the whistler waves which conventionally arise in Hall plasmas. However, whistler waves stable to the leading order, exhibit the instability in the next order approximation in which the non-local effects are already become significant. The reason is that due to the finite gradients of the steady state equilibrium variables the resulting eigen-frequencies are strong functions of space, a fact that renders the validity of that so called local approximation doubtful. The overall result of the present study is that in spite of stability of a Hall equilibrium disc to leading order in high-frequency, the whistler waves can be amplified in the next order approximation. Thus, amplification of the whistler mode is entirely due to density stratification of the plasma at the scales of the Keplerian discs thickness, while the amplification factor is independent of the magnetic field. The amplification factor is independent also of the choice of arbitrary trial functions and boundary function which reflect the uncertainty of the steady-state equilibrium solution. It was found that rotation and compressibility effects are negligibly small for both propagation and amplification of whistler waves.

It is worthwhile to mention that the condition of zero equilibrium-density ($n_v = 0$) at the horizontal disc edges ($\zeta = \pm 1$) is a direct consequence of the sharp boundary disc that has been employed in the present investigation. Since the amplification factor is a power function of the number density, i.e., $A(\zeta) = 1/\sqrt{n_v(r,\zeta)/n_{vm}(r)}$, the perturbations are amplified as a power function of the distance from the mid-plane to the disc edge in a finite thickness large disc with $n_v/n_{vm} = (1-\zeta^2)^{1/(\gamma-1)}$. In fact, astrophysical discs have no well defined boundaries, the disc boundary is not sharp, and the vacuum boundary conditions posed at a finite height are rather a good approximation than an exact model. According to a more realistic model that assumes a diffused disc with exponentially decreasing density [see e.g. Balbus & Hawley (1991) for thin isothermal discs, where $n_v/n_{vm} \sim \exp(-\sigma\zeta^2)$, $\sigma \sim 1$], the density has a small but finite value at the effective boundary of the disc. Thus, assuming on a phenomenological base the exponentially decreasing density ($n_v/n_{vm} \sim \exp(-\sigma\zeta)$, $\sigma \sim 1$) this is equivalent to exponential amplification factor $A(\zeta) = 1/\sqrt{n_v(r,\zeta)/n_{vm}(r)} \sim \exp(\sigma\zeta/2)$. Note also that the exponential spatial amplification of the waves is closely related to the notion of convective instability [e.g. Stix (1992)]. As distinct from an absolute instability in which an initial disturbance produces a response that grows exponentially with time at arbitrary spatial point, in a convective instability, the response stops to grow after the disturbance has passed any fixed spatial point. The distinction between absolute and convective instability is rather conditional, the response to absolute instability can appear as a convective instability and otherwise depending on the chosen coordinate frame. Thus, spatially amplified time-periodic disturbances wave will increase in time in the frame traveling with the disturbance. In particular, the growth rate in time instability may be estimated. Using an estimation for group velocity of whistler waves $C_g \sim \omega_H^{1/2}$ yields the dimensionless growth rate coefficient of the order of $\omega_H^{1/2}\sigma \sim \omega_H^{1/2}$ in the scale of the Keplerian rotation time. Hence, the corresponding characteristic growth time will be much less than the Keplerian rotation time for high frequencies $\omega_H$.

Finally note that the whistler waves expand rather through almost the whole Keplerian portion of the disc, and have enough time for multiple reflections from the horizontal edges during a single Keplerian rotation time. The possible effect of the wave amplification due to its multiple reflections from effective horizontal edges ignored in the present study was considered in Liverts & Mond (2009), where the existence and role of the edges were discussed in the framework of the MHD plasma model.

**Appendix A. EXPLICIT EQUILIBRIUM SOLUTIONS FOR SMALL INVERSE HALL PARAMETER**

Equilibrium magnetic field and density are governed by the choice of the arbitrary functions $\varphi=\varphi(\psi)$ and $I = I(\psi)$ in Eqs. (15) and (19). According to Shtemler et al. (2009) linear trial functions may be utilized in order to qualitatively describe equilibrium state:

$$I(\psi)= I_0 + I_1\psi, \qquad \varphi(\psi)= \varphi_0+\varphi_1 \psi, \qquad ( dI/d\psi = I_1 \equiv const,\ d\varphi/d\psi = \varphi_1 \equiv const ). \qquad (A1)$$

The constant $I_0$ is the reference total current through the disc that is localized along the disc axis, which describes the central body effect. Such a modeling appears to be crude at first sight, but it is quite efficient assuming the structure of the magnetic field across a thin disc is rather simple.

It is conjectured that the main features of the equilibrium are sufficiently general since relations (A1) may be viewed as a best fit linear approximations. Since solution of equilibrium problem is invariant with respect to the simultaneous change of $I_0$, $I_1$ by $-I_0$, $-I_1$, further analysis may be restricted to positive values of $I_0$. Furthermore, the characteristic magnetic field $B_*$ is chosen here such that the dimensionless total current through the disc central part is $I_0 = 1$. Such a normalizing is convenient for the classification of the formal equilibrium solutions presented in Shtemler et al. (2009). Indeed, different limits in inverse Hall parameter $\delta_H$ can be considered depending on the relative values of the magnetic dipole strength in the boundary function $\Gamma = -M_d/r$, and of the parameters in the GS equation (15), such as the deviations, $I_1$ and $\varphi_1$, of the total current and electric potential from the reference values $I_0 = 1$ and, without loss of generality, $\varphi_0 = 0$.

In the adiabatic case adopted here $\gamma = 5/3$ ($\nu \approx 0.1$), the inverse Hall parameter $\delta_H = \nu/\pi_H$ varies from $\delta_H = 0$ to $\delta_H \approx 0.1$ for the Hall parameter ranged from $\pi_H = \infty$ to $\pi_H = 1$. Thus, asymptotic expansions in $\delta_H \ll 1$ for equilibrium density, pressure, velocity and constants of the trial functions provide explicit solutions of the problem Shtemler et al. (2009):

$$n_\nu = n^{(0,0)} + \delta_H n^{(0,1)} + ..., \quad P_\nu = P^{(0,0)} + \delta_H P^{(0,1)} + ..., \quad V_\theta = V_\theta^{(0,0)} + \delta_H V_\theta^{(0,1)} + ...,$$

$$I_1 = I_1^{(0,0)} + \delta_H I_1^{(0,1)} + ..., \quad \varphi_1 = \varphi_1^{(0,0)} + \delta_H \varphi_1^{(0,1)} + ..., \quad M_d = M_d^{(0,0)} + \delta_H M_d^{(0,1)} + .... \quad (A2)$$

Here the first and second superscripts correspond to the orders in $\varepsilon$ and $\delta_H$, respectively. The scaling in $\delta_H$ is applied to the unknown functions that are already scaled with $\varepsilon$. At small $\delta_H$ there are two families of the equilibrium discs which correspond to small and large discs, correspondingly.

Small-radius discs (first family of the equilibrium solutions) which occur for large deviations $I_1 \sim \delta_H^0$ from a reference total current $I_0 = 1$, and to magnetic dipole strengths $M_d \lesssim \delta_H^0$, or, with no generality lost:

$$I_1 \approx I_1^{(0,0)}, \qquad \varphi_1 \approx \varphi_1^{(0,0)}, \qquad M_d \approx M_d^{(0,0)}. \tag{A3}$$

Substituting estimations (A3) into the density distribution (19) and using them in the GS Eq. (15) supplemented by boundary conditions (16), (17) yield the following explicit expressions for density and magnetic flux:

$$n_v \approx \left\{ \frac{2}{\beta_H} [\psi - \Gamma(r)][\Omega(r) + \varphi] + \frac{1-\zeta^2}{r^3} \right\}^{3/2}, \quad \psi \approx -\frac{I_0}{I_1} + \frac{I_0 + I_1 \Gamma(r)}{I_1} \frac{\cos(I_1 \zeta)}{\cos(I_1)}, \quad rB_\theta \equiv I(\psi), \tag{A4}$$

where $\Omega(r) = r^{-3/2}$, $\Gamma(r) = -M_d/r$. The first family of equilibria is characterized by a finite disc radius that is determined by the condition of non-negativity of the expression for the number density in (A4). In the limiting case of large Hall plasma beta $\beta_H = \beta \pi_H \gg 1$, first family of the equilibrium solutions for small-radius discs is rather degenerated to the equilibrium solutions for large radius discs since the disc radius $\sim \beta_H^{1/3}$ infinitely rises with $\beta_H$. Since in the Hall regime $\pi_H \gtrsim 1$, small-radius discs may exist at small plasma beta $\beta$ such that $\beta_H = \beta \pi_H \lesssim 1$.

Large radius discs (second family of the equilibrium solutions) correspond to small deviations ($I_1 \lesssim \delta_H$) from the reference total current $I_0 = 1$. In that case the GS Eq. (15) together with the edge boundary conditions yield

$$I_1 \approx \delta_H I_1^{(0,1)}, \quad \varphi_1 \approx \varphi_1^{(0,0)}. \tag{A5}$$

As a result, to leading order in $\delta_H$ Eq. (19) for the number density is reduced to a pure hydrodynamic form, while the equilibrium GS problem (15)-(18) yields:

$$n_v(r,\zeta) \approx \left[\frac{1-\zeta^2}{r^3}\right]^{3/2}, \quad \psi(r,\zeta) - \Gamma(r) = \delta_H \psi^{(0,1)}(r,\zeta), \quad rB_\theta = I(\psi) \approx 1, \tag{A6}$$

where

$$\psi^{(0,1)} = I_1^{(0,1)} \frac{1-\zeta^2}{2} + \frac{\Omega(r) + \varphi_1^{(0,0)}}{8r^{5/2}} X(\zeta), \quad X(\zeta) = \sqrt{1-\zeta^2}\left[1-\zeta^2 + \frac{2}{5}(1-\zeta^2)^2 - 3\frac{\zeta \arcsin \zeta - \pi/2}{\sqrt{1-\zeta^2}} - 3\right].$$

Different limits of (A6) can be additionally considered now. For instance, the magnetic dipole $M_d$ may be of the order of or much larger/less than $\delta_H$. Thus, Eq. (A6) describes two different classes of solutions corresponding to (i) weak and (ii) strong magnetic-dipoles.

(i) The first class of large equilibrium discs (i.e. of the second family of the equilibrium solutions) corresponds to a weak magnetic-dipole strength:

$$M_d = \delta_H M_d^{(0,1)}, (\Gamma(r) = \delta_H \Gamma^{(0,1)}(r)), n_v(r,\zeta) \approx \left[\frac{1-\zeta^2}{r^3}\right]^{3/2}, \psi \approx \delta_H [\Gamma^{(0,1)}(r) + \psi^{(0,1)}(r,\zeta)], rB_\theta = I(\psi) \approx 1.$$

$$\tag{A7}$$

(ii) The second class of large equilibrium discs (i.e. of the second family of the equilibrium solutions) corresponds to a strong magnetic-dipole strength:

$$M_d = \delta_H^0 M_d^{(0,0)}, \quad (\Gamma(r) \sim \delta_H^0), n_v(r,\zeta) \approx \left[\frac{1-\zeta^2}{r^3}\right]^{3/2}, \psi(r,\zeta) \approx \Gamma(r), \quad rB_\theta = I(\psi) \approx 1. \tag{A8}$$

A specific flux function $\psi(r,\zeta) \approx \Gamma(r)$ in (A8) neglects the input of the Hall effects into equilibrium solution, but it is governed by the magnetic dipole value due to the central body effect.